\documentclass[11pt,twoside]{article}
\usepackage{asp2010}

\resetcounters

\begin{document}

\title{3D hydrodynamical simulations of evolved stars and observations
of stellar surfaces}
\author{A.~Chiavassa$^1$ and B.~Freytag$^2$
\affil{$^1$Laboratoire Lagrange, UMR 7293, CNRS, Observatoire de la C\^ote d'Azur, Universit\'e de Nice Sophia-Antipolis, Nice, France}
\affil{$^2$Astronomical Observatory, Uppsala University, Regementsvagen 1, Box 515, SE-75120 Uppsala, Sweden}}

\begin{abstract}
Evolved stars are among the largest and brightest stars and they are ideal targets for the new generation of sensitive, high resolution instrumentation that provides spectrophotometric, interferometric, astrometric, and imaging observables. The interpretation of the complex stellar surface images requires numerical simulations of stellar convection that take into account multi-dimensional time-dependent radiation hydrodynamics with realistic input physics. We show how the evolved star simulations are obtained using the radiative hydrodynamics code CO5BOLD and how the accurate observables are computed with the post-processing radiative transfer code {\sc Optim3D}. The synergy between observations and theoretical work is supported by a proper and quantitative analysis using these simulations, and by strong constraints from the observational side.
\end{abstract}

\section{Introduction}

Evolved stars are characterized by low surface gravity (lower than $\sim\log g=1.0$) and their atmosphere is unstable against convection in deep layers with poorly defined boundaries. This instability goes even deeper inwards for most of the distance to the stellar center with, as an extreme case, the Asymptotic Giant Branch (AGB) stars. Already \cite{1975ApJ...195..137S} estimated that Òonly a modest number of them exists at any one time on the entire surfaceÓ of Red SuperGiant (RSG) stars, based on the assumption that the horizontal size of a granule scales with a characteristic vertical length as the pressure scale height. However, only recently highly accurate interferometric and satellite observations managed to image the stellar surface \citep[e.g., ][and Paladini's contribution to this conference]{1996ApJ...463L..29G,1997ApJ...482L.175K,2000MNRAS.315..635Y,2009ApJ...707..632L,2009A&A...508..923H,2010A&A...511A..51C,2011A&A...529A.163O,2014A&A...568A..17O}. 
The low effective temperature (lower than $\sim$ 4000K) is a source of spectra so crowded with lines from atoms and molecules that, in particular in the optical region where electronic transitions of some abundant molecule occur, there is little hope to see the continuum forming region, even at high spectral resolution. In addition to this, many molecular data needed when a model atmosphere is computed or used for abundance analysis, is still not accurately known. 
Ultimately, the derivation of their stellar parameters such as effective temperature and surface gravity is not trivial, and, in particular, the surface gravity is still highly uncertain.\\
The understanding of the dynamical convective pattern of evolved stars is then crucial for the comphrension of the physics of these stars that contribute extensively to the chemical and dusty enrichment of galaxies.

\section{Three-dimensional modelling and post-processing radiative transfer}

The effects of convection and non-radial waves can be represented
by numerical multi-dimensional time-dependent radiation
hydrodynamics (RHD) simulations with realistic input
physics. The results presented are based on the computations carried out with {\sc CO5BOLD} \citep[][and Freytag's contribution to this conference]{2012JCoPh.231..919F,2008A&A...483..571F,2002AN....323..213F}. {\sc CO5BOLD} solves the coupled equations of compressible hydrodynamics and non-local radiation transport. The hydrodynamics module is based on a finite-volume approach and uses Riemann solver of Roe-type \citep{1986AnRFM..18..337R}. The equation of state uses pre-tabulated values as functions
of density and internal energy $\left(\rho,e_i\rightarrow P,\Gamma_1,T,s\right)$. It accounts for H{\sc{I}}, H{\sc{II}}, H$_{2}$, He{\sc{I}}, He{\sc{II}}, He{\sc{III}} and a representative metal for any prescribed chemical composition.\\
In the case of RSG and AGB stars, the \textsc{star-in-a-box} geometry is applied and the computational domain is a
cubic grid equidistant in all directions; the same
open boundary condition is employed for all sides of the computational box. The radiation transport for the simulations of evolved stars employs a short-characteristics
method, and, to account for the short radiative time scale, several
(typically 6 to 9) radiative sub$-$steps are performed per
global step. The simulations can be computed: (i) either using a gray frequency dependance of the radiation field, which ignores the frequency dependence, based on Rosseland mean opacities calculated merging high-temperature OPAL \citep{1992ApJ...397..717I} data and low-temperature
PHOENIX \citep{1997ApJ...483..390H} at around 12\ 000K; (ii) or using a multi-group scheme \citep{1994A&A...284..105L,2004A&A...421..741V}, where the frequencies that reach monochromatic optical depth unity within a certain depth range of the model atmosphere will be put into one frequency group. The RHD simulations employing the latter method have typically five wavelengths groups sorted according to the run of the monochromatic optical depth in a corresponding MARCS \citep{2008A&A...486..951G} 1D model with a smooth transition to the Rosseland mean (OPAL opacities) in the optically thick regime.\\

\articlefigure{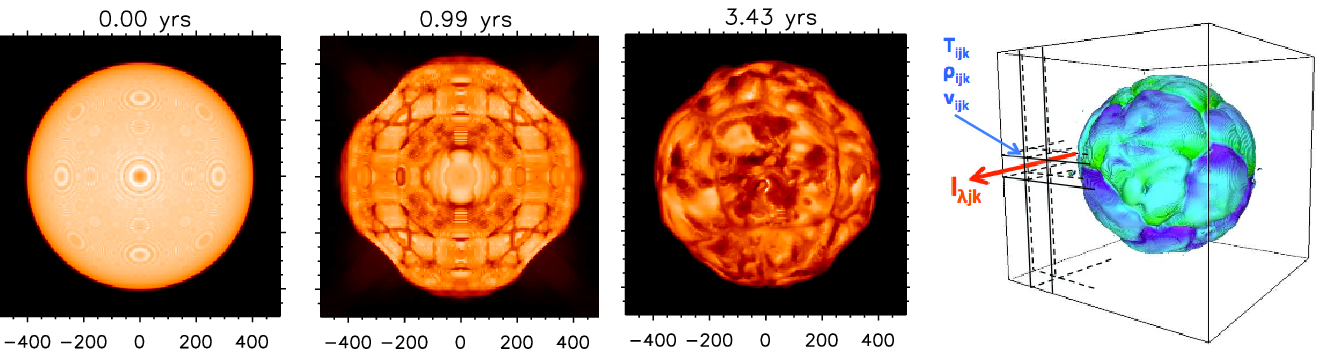}{fig1}{\emph{First, second and third panels: }gray intensity on one side of the computational cube from the initial sequence of RSG simulation. The axes are in solar radii \citep{2011A&A...535A..22C}. \emph{Fourth panel:} Three-dimensional volume rendering of the temperature in one RSG simulation, together with the numerical grid with the variable temperature ($T$), density ($\rho$), and velocity ($v$) read by {\sc Optim3D} and used to compute the outgoing monochromatic intensity ($I_\lambda$).}

In {\sc CO5BOLD}, the most important parameters \citep{2011A&A...535A..22C} that determine the type of the simulated star are:
\begin{itemize}
\checklistitemize
\item the input luminosity into the core
\item the stellar mass that enters in the equation for the gravitational potential
\item the abundances that are used to create the tables for the equation-of-state and the opacities.
\end{itemize}

The initial model is produced starting from a sphere in hydrostatic equilibrium with a weak velocity field inherited from a previous model with different stellar parameters (Fig.~\ref{fig1}; first, second, and third panels). After some time, the limb-darkened surface without any convective signature appears but with some regular patterns due to the numerical grid. The central spot, quite evident at the beginning of the simulation, vanishes completely when convection becomes strong. After several years of stellar time, a regular pattern of small-scale convection cells develops and, after cells merge the average structures, it becomes big and the regularity (due to the Cartesian grid)
is lost. The intensity contrast grows with time.

Once the RHD simulation is relaxed, the snapshots are used for detailed post-processing treatment to extract interferometric, spectrophotometric, astrometric, and imaging observables that are compared to the observations to tackle different astrophysical problems, as well as, to constrain the simulations. For this purpose, we use the 3D pure-LTE radiative transfer code {\sc Optim3D} \citep{2009A&A...506.1351C} to compute synthetic spectra and intensity maps from
the snapshots of the RHD simulations. The code takes into account the
Doppler shifts due to convective motions. The radiative
transfer equation is solved monochromatically using pre-tabulated extinction coefficients as a function of temperature, density, and
wavelength. The lookup tables were computed using the same extensive atomic and molecular opacity data as the latest generation of
MARCS models \citep{2008A&A...486..951G}. We assumed a zero
micro-turbulence since the velocity fields inherent in 3D models
are expected to self-consistently and adequately account for
non-thermal Doppler broadening of spectral lines. The temperature and density ranges
spanned by the tables are optimized for the values encountered in the RHD simulations. \\
{\sc Optim3D} computes the emerging intensities for vertical rays cast through the
computational box, for all required wavelengths (Fig.~\ref{fig1}, fourth panel).\\

With the synergy between {\sc CO5BOLD} and {\sc Optim3D}, we produced a set of observables covering all the wavelengths from optical to far infrared. We present the results in the remaining sections of this proceedings.

\newpage

\section{Simulations of RSG and AGB stars}

RHD simulations of evolved stars show a very heterogeneous surface caused by the dynamical granulation. The emerging intensity is related to layers where waves and shocks dominate together with the variation in opacity through the atmosphere. Small-amplitude acoustic waves are produced in the convective
envelope by non-stationary convective flows with
significant Mach numbers (e.g., 5 or even larger). These waves can
travel outward in the convective envelope and even into the
convectively stable atmosphere where they are compressed and
amplified (due to the lower temperature and sound speed) and further
amplified (due to the lower density). Here, they turn into shocks
giving rise to a dynamical pressure larger than the gas pressure \citep[typically a factor 5, see e.g.][the bottom row of Fig.~2 in the paper]{2011A&A...535A..22C}.\\
In summary, RHD simulations pulsate by themselves and do not have any dynamic boundary condition, the hydrodynamical equations
include the advection of momentum, which, after averaging over
space and time, gives the dynamical pressure. AGB models show much extend and varying structures in the near-surface than RSG, and, convective velocities in RSGs are too low to reach escape velocity and (fully) contribute to the mass-loss mechanism. RSG and AGB simulations are both characterized by  large convective cells and strong shocks, however, AGBs have in general even larger scales with shocks pushing the mass much further out. Typical RHD simulations of RSGs and AGBs have the stellar parameters reported in Table~\ref{simus}.\\

\begin{table}[!ht]
\caption{Typical stellar parameters for RHD simulations of RSG \citep{2011A&A...535A..22C} and AGB \citep[][and Freytag's contribution to this conference]{2008A&A...483..571F} stars.}
\label{simus}  
\smallskip
\begin{center}
{\small
\begin{tabular}{cccccccc}
\tableline
\noalign{\smallskip}

Simulation & Numerical & $M_{\mathrm{pot}}$ & $M_{\mathrm{env}}$ & $L$ &  $T_{\rm{eff}}$ & $R_{\star}$ &  $\log g$   \\
                  & resolution & $\!\!\!$[$M_\odot$]$\!$ & $\!\!\!$[$M_\odot$]$\!$ & $\!\!\!$[$L_\odot$]$\!$ & $\!\!\!$$[\rm{K}]$$\!$ & $\!\!\!$[$R_\odot$]$\!$ &  $\!\!\!$[cgs]$\!$  \\

\noalign{\smallskip}
\tableline
\noalign{\smallskip}
RSG & $401^3$ &  12   & 3 &  90000 & 3500 & 840 & $-$0.33  \\
AGB & $401^3$ & 1 & 0.186 & 7000 & 2500 & 430 & $-$0.83  \\
\noalign{\smallskip}
\tableline
\end{tabular}
}
\end{center}
\end{table}

Monochromatic intensity maps from RHD simulations are characterized by strong wavelength dependence that changes the appearance of the stellar surface from the optical to the infrared (Fig.~\ref{fig2}). This behavior is particularly important for the AGBs (Fig.~\ref{fig5}). Moreover, also the temporal time scales of the granulation pattern are different with respect to the spectral range probed. An example is reported in Fig.~\ref{fig2}, where the simulation of a RSG star is characterized by two principal characteristic time scales linked directly to the stellar dynamical effects:

\begin{itemize}
\item the surface of the RSGs is covered by a few large convective cells with a size of $\approx60\%$ of the stellar radius (top row of Fig.~\ref{fig2}) that evolve on a time scale of years \citep{2009A&A...506.1351C}. This is visible in the infrared, and particularly in the H band where the H$^-$ continuous opacity minimum occurs and consequently the continuum-forming region is more evident;
\item in the optical region \citep[bottom row of Fig.~\ref{fig2}, ][]{2011A&A...528A.120C}, short-lived (a few weeks to a few months) small-scale ($<10\%$ of the stellar radius) structures appear. They result from the opacity contribution and dynamics at optical depths smaller than 1 (i.e., further up in the atmosphere with respect to the continuum-forming region), as well as from the higher sensitivity of the blackbody radiation to the temperature inhomogeneities. It must be noted that also the numerical resolution of the simulation plays a role for the size
and number of these small structures.
\end{itemize}

\articlefigure{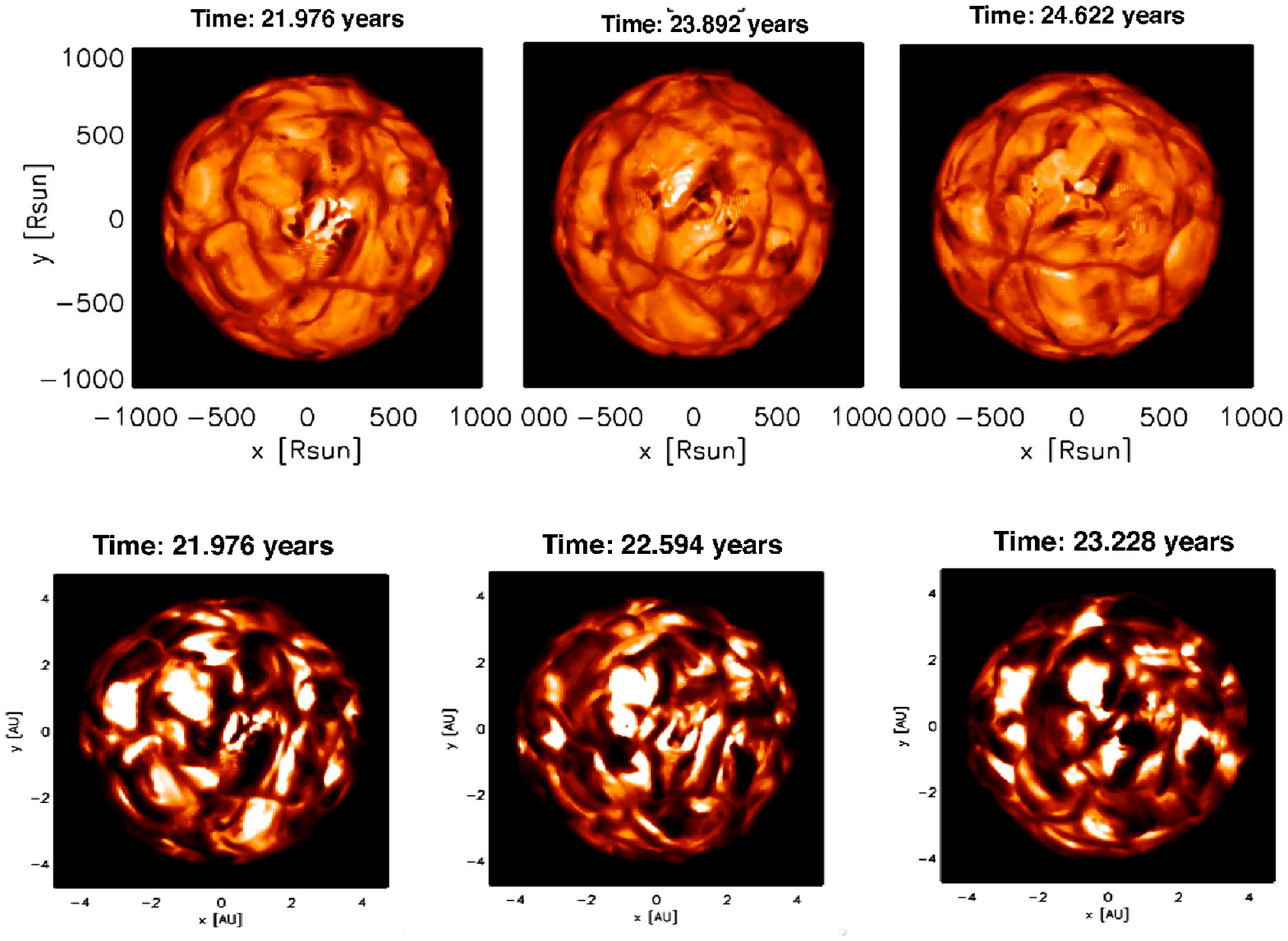}{fig2}{Synthetic maps (linear intensity) of a RSG simulation with stellar parameters of Table~\ref{simus} at different stellar time in the H band \citep[top row,][]{2009A&A...506.1351C} and in the optical region \citep[bottom row,][]{2011A&A...528A.120C}.}

In the following sections, we present results mostly concerning RGSs. However, the same analysis and qualitatively similar results have to be expected for AGB stars, but in a more extreme way. 

\section{Spectroscopy - low resolution}

\begin{figure*}
   \centering
   \begin{tabular}{ccc}
           \plotone[width=0.98\hsize]{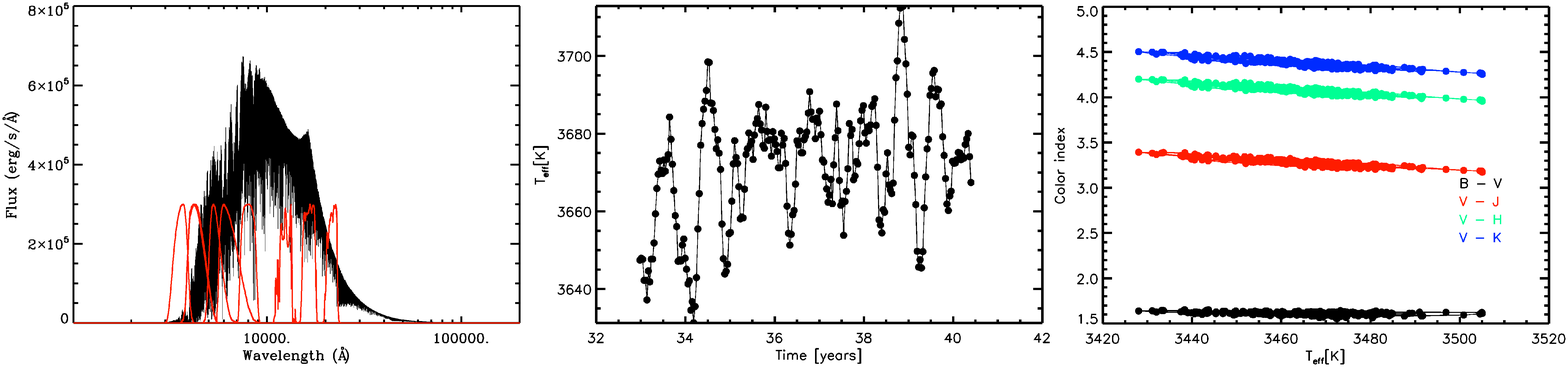} 
              \end{tabular}
      \caption{\emph{Left panel: } spectral energy distribution for one snapshot of a RSG simulation with stellar parameters of Table~\ref{simus}. A billion of molecular and atomic line transitions are included in the calculation. The red curves correspond to different photometric filters. \emph{Central panel: } effective temperature computed from synthetic spectra of different simulation snapshots as a function of stellar time. \emph{Right panel: }different color index as a function of the corresponding effective temperatures.
           }
        \label{fig3}
   \end{figure*}

It is worthwhile to analyze separately the results of RSG simulations for low and high resolution spectra. An important aspect of RHD simulations is the temporal variability. The vigorous convective motions and inhomogeneities cause large fluctuations in the spectra that will affect {\sc Gaia} spectrophotometric measurements up to 0.28 mag in the blue photometric range and 0.15 mag in the red filter \citep[][Section~3.2]{2011A&A...528A.120C}. Moreover, also the stellar parameters, such as effective temperature, and photometric colors fluctuate. Fig.~\ref{fig3} shows an example of synthetic spectrum of a RSG (left panel) for which the effective temperature ($T_{\rm{eff}}$) has been computed using Stefan-Boltzmann law as $T_{\rm{eff}} = \left\{\left[\int_{\lambda_1}^{\lambda_2}f\left(\lambda\right)d\lambda\right]/\sigma\right\}^{0.25}$ where $\lambda_1$=1010 \AA\ and $\lambda_2$=199\ 960 \AA\ , $f\left(\lambda\right)$ is the synthetic flux, and $\sigma$ is the Stefan-Boltzmann constant (the typical temperature and radius are reported in Table~\ref{simus}). The temporal evolution of the effective temperature is shown in central panel of Fig.~\ref{fig3} with maximum and minimum fluctuations ranging from $\sim$3430 to $\sim$3505 K and an average value of $\sim$3470K, over a period of about eight stellar years. Broad-band colors are usually tightly correlated with the stellar effective temperature, although metallicity ([Fe/H]) and (to a lesser extent) surface gravity (log g) may play a role. Fig.~\ref{fig3} (right panel) shows different color-color plots as a function of the corresponding effective temperature. The colors appear largely scattered and, also in this case, the principal reasons are the vigorous convective motions and the surface inhomogeneities. \\
Stellar parameters determination resides on the dynamical effects of the atmosphere and are not constant with respect to temporal evolution. The same behavior is expected for AGBs, but even in a more extreme way.

\section{Spectroscopy - high resolution}

RHD simulations provide a self-consistent ab-initio description of the non-thermal velocity field generated by convection, shock waves, and overshoot that manifests itself in spectral line shifts and changes in the equivalent width (Fig.~\ref{fig4}, top left). The shape of the optical Ti I line at 6261.11 \AA\ , taken as an example here, constitutes of more than one velocity component that contributes through the different atmospheric layers where the line forms. As a consequence, the line bisector\footnote{It is the locus of the midpoints of the line. A symmetric profile has a straight vertical bisector, while the "C"-shaped line bisector reveals asymmetries.} is not straight and span values up to 5 km/s on a temporal scale of few weeks (Fig.~\ref{fig4}, top and bottom right panels). \cite{2008AJ....135.1450G} showed that the prototypical RSG star $\alpha$~Ori displays a characteristic reverse C-shape signature on the Ti I line with line shifts up to $\sim$ 9 km/s (Fig.~\ref{fig4}, bottom left panel). RHD simulation of RSGs are in qualitatively good agreement with these observations in term of the line bisector morphology and evolution as well as the spanned velocities. There is, however, a difference in the shifted velocities which appear blue-shifted in the observations and red-shifted in the simulation. This issue is under investigation. 
As the vigorous convection is prominent in the emerging flux, the radial velocity measurements for evolved stars are very complex and need a sufficiently high spectral resolution to possibly disentangle all the sources of macro-turbulence.  \\

RHD simulations are also used to tune hydrostatic models that only use empirical calibrations, such as micro- and macro-turbulence velocities, to approximate the turbulent flow. \cite{2011A&A...535A..22C} provided a calibration of these parameters for 1D models using RGSs simulations (micro-turbulence ranging from 1.28 to 1.45 km/s) and they argued that the depth-independent micro-turbulence assumption in 1D models is a fairly good approximation. They also found that the 3D-1D corrections of the resulting iron abundances are quite small (at least for the simulations and iron lines considered in that work). Finally, the authors assessed that there is no clear distinction between the different macroturbulent profiles chosen in 1D models to fit 3D synthetic lines.

\articlefigure{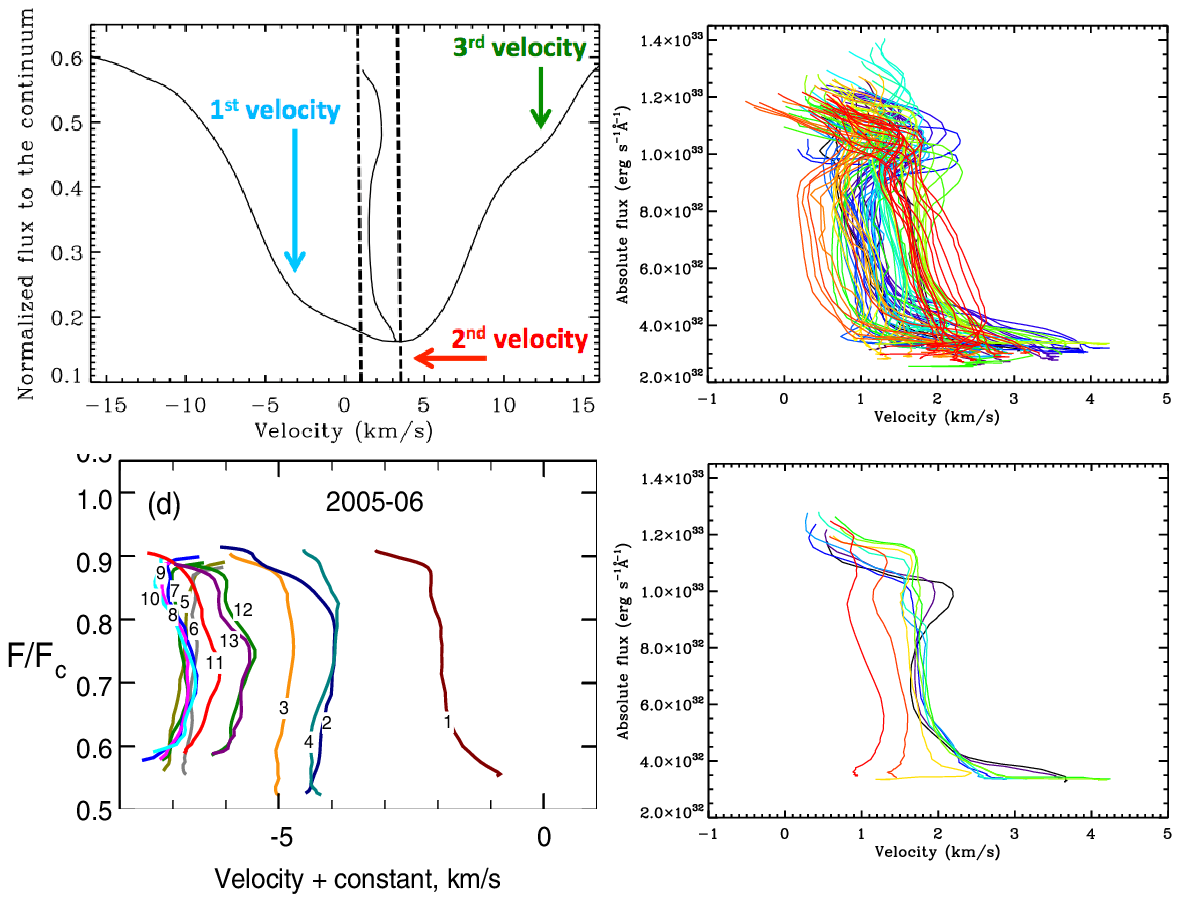}{fig4}{\emph{Top left panel: }synthetic spectrum of Ti I line at 6261.11 \AA\ for one snapshot of a RSG simulation with stellar parameter of Table~\ref{simus}. The vertical dashed line shows the spanned velocities of the line bisector. The different arrow and colors displays the positions of different velocity components. \emph{Top right panel: } temporal series of Ti I line bisectors derived from a RSG simulation. \emph{Bottom left panel: } bisectors of Ti I line observed for the RSG star $\alpha$~Ori by \cite{2008AJ....135.1450G}. \emph{Bottom right panel: }enlargement of the top right panel for few snapshots.}

\begin{figure*}
   \centering
   \begin{tabular}{c}
      \includegraphics[width=0.98\hsize]{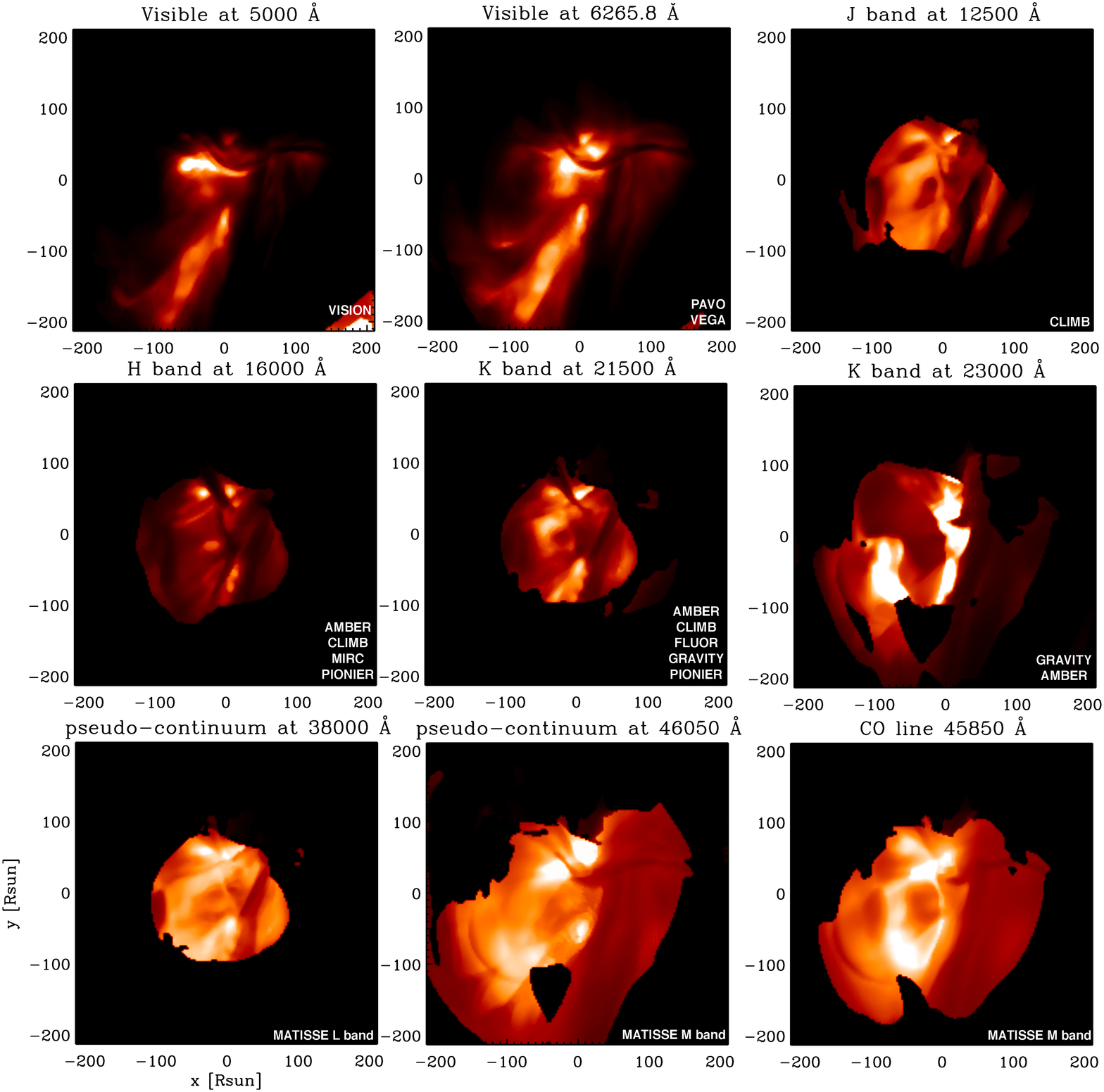} 
              \end{tabular}
      \caption{Wavelength series of synthetic images (linear intensity) computed for a simulation of an AGB star with the stellar parameters of Table~\ref{simus}. In lower right corner of each panel, the name of an interferometric instrument (in alphabetic order) that probe the corresponding wavelengths.
           }
        \label{fig5}
   \end{figure*}

\section{Interferometry}

Interferometry in crucial for evolved stars because allows the direct detection and characterization of the convective pattern related to the surface dynamics. Two main observables are used in interferometry: the visibility and the closure phases. The combination of both, plus a good enough coverage of the Fourier plane, contributes to the image reconstruction of the observed targets. Visibilities measure the surface contrast of the source and are primarily used to determine the fundamental stellar parameters and the limb-darkening. Closure phases combine the phase information from three (or more) telescopes and provide direct information on the morphology of the source. For a correct interpretation of the observations, it is necessary to simultaneously explain both observables with the same model.

\cite{2010A&A...515A..12C} detected and measured the characteristic sizes of convective cells on the RSG star $\alpha$~Ori using visibility measurements in the infrared. Moreover, they managed to explain the observation of $\alpha$~Ori from the optical to the infrared region using RHD simulations, showing that its surface is covered by a granulation pattern that, in the H and K bands, shows structures with small to medium scale granules (5-15 mas, while the size of the star at these wavelengths is $\sim$44 mas) and a large convective cell ($\sim$30 mas). Another result concerns the first reconstructed images with AMBER/VLTI of the massive evolved star VX~Sgr \citep{2010A&A...511A..51C}. The authors used RHD simulations of RSG and AGB to probe the presence of large convective cells on its surface.

Interferometric predictions for evolved stars are used to observe their convective pattern ranging from the optical to the far infrared and to evaluate its effect on the observables. In this context, Fig.~\ref{fig5} displays the surface of a simulated AGB with stellar parameters of Table~\ref{simus}. The wavelength dependence is striking, but it is also important to note that going from the infrared to the optical, there is a relevant increase of the intensity contrast\footnote{In average, the brightest areas exhibit an intensity $\sim$50 times or larger than the dark ones in the optical and up to $\sim$10 times in the infrared. Probing very narrow wavelength filters close to particular spectral line centers may increase these values.} as well as the number and complexity of surface structures. In addition to this, as reported in previous Sections, also the temporal evolution (at different wavelengths) in a key point in the understanding of stellar dynamics. To tackle all the different astrophysical problems related to the evolved stars, today and future interferometers have to challenge the combination of high spectral and spatial resolution as well as the possibility to observe (i.e., to image) the same object several time in a year. Additionally, the "relatively" new spectral window in the visible and J band must be developed further.

\section{Summary}

The cooperation between the radiation hydrodynamical simulations computed with {\sc CO5BOLD} coupled with the post-processing radiative transfer code {\sc Optim3D} is essential to a proper quantitative analysis of interferometric, spectrophotometric, astrometric, and imaging observations of evolved stars. In addition to this, the advent of highly accurate observations largely helps the development of the theoretical work with strong constraints.

\bibliography{chiavassa}

\begin{thebibliography}{}
\expandafter\ifx\csname natexlab\endcsname\relax\def\natexlab#1{#1}\fi
\expandafter\ifx\csname url\endcsname\relax
  \def\url#1{\texttt{#1}}\fi
\expandafter\ifx\csname urlprefix\endcsname\relax\def\urlprefix{URL }\fi
\providecommand{\eprint}[2][]{\url{#2}}

\bibitem[{{Chiavassa} et~al.(2011{\natexlab{a}}){Chiavassa}, {Freytag},
  {Masseron}, \& {Plez}}]{2011A&A...535A..22C}
{Chiavassa}, A., {Freytag}, B., {Masseron}, T., \& {Plez}, B.
  2011{\natexlab{a}}, \aap, 535, A22. \eprint{1109.3619}

\bibitem[{{Chiavassa} et~al.(2010{\natexlab{a}}){Chiavassa}, {Haubois},
  {Young}, {Plez}, {Josselin}, {Perrin}, \& {Freytag}}]{2010A&A...515A..12C}
{Chiavassa}, A., {Haubois}, X., {Young}, J.~S., {Plez}, B., {Josselin}, E.,
  {Perrin}, G., \& {Freytag}, B. 2010{\natexlab{a}}, \aap, 515, A12.
  \eprint{1003.1407}

\bibitem[{{Chiavassa} et~al.(2010{\natexlab{b}}){Chiavassa}, {Lacour},
  {Millour}, {Driebe}, {Wittkowski}, {Plez}, {Thi{\'e}baut}, {Josselin},
  {Freytag}, {Scholz}, \& {Haubois}}]{2010A&A...511A..51C}
{Chiavassa}, A., {Lacour}, S., {Millour}, F., {Driebe}, T., {Wittkowski}, M.,
  {Plez}, B., {Thi{\'e}baut}, E., {Josselin}, E., {Freytag}, B., {Scholz}, M.,
  \& {Haubois}, X. 2010{\natexlab{b}}, \aap, 511, A51. \eprint{0911.4422}

\bibitem[{{Chiavassa} et~al.(2011{\natexlab{b}}){Chiavassa}, {Pasquato},
  {Jorissen}, {Sacuto}, {Babusiaux}, {Freytag}, {Ludwig}, {Cruzal{\`e}bes},
  {Rabbia}, {Spang}, \& {Chesneau}}]{2011A&A...528A.120C}
{Chiavassa}, A., {Pasquato}, E., {Jorissen}, A., {Sacuto}, S., {Babusiaux}, C.,
  {Freytag}, B., {Ludwig}, H.-G., {Cruzal{\`e}bes}, P., {Rabbia}, Y., {Spang},
  A., \& {Chesneau}, O. 2011{\natexlab{b}}, \aap, 528, A120. \eprint{1012.5234}

\bibitem[{{Chiavassa} et~al.(2009){Chiavassa}, {Plez}, {Josselin}, \&
  {Freytag}}]{2009A&A...506.1351C}
{Chiavassa}, A., {Plez}, B., {Josselin}, E., \& {Freytag}, B. 2009, \aap, 506,
  1351. \eprint{0907.1860}

\bibitem[{{Freytag} \& {H{\"o}fner}(2008)}]{2008A&A...483..571F}
{Freytag}, B., \& {H{\"o}fner}, S. 2008, \aap, 483, 571

\bibitem[{{Freytag} et~al.(2002){Freytag}, {Steffen}, \&
  {Dorch}}]{2002AN....323..213F}
{Freytag}, B., {Steffen}, M., \& {Dorch}, B. 2002, Astronomische Nachrichten,
  323, 213

\bibitem[{{Freytag} et~al.(2012){Freytag}, {Steffen}, {Ludwig},
  {Wedemeyer-B{\"o}hm}, {Schaffenberger}, \& {Steiner}}]{2012JCoPh.231..919F}
{Freytag}, B., {Steffen}, M., {Ludwig}, H.-G., {Wedemeyer-B{\"o}hm}, S.,
  {Schaffenberger}, W., \& {Steiner}, O. 2012, Journal of Computational
  Physics, 231, 919. \eprint{1110.6844}

\bibitem[{{Gilliland} \& {Dupree}(1996)}]{1996ApJ...463L..29G}
{Gilliland}, R.~L., \& {Dupree}, A.~K. 1996, \apjl, 463, L29

\bibitem[{{Gray}(2008)}]{2008AJ....135.1450G}
{Gray}, D.~F. 2008, \aj, 135, 1450

\bibitem[{{Gustafsson} et~al.(2008){Gustafsson}, {Edvardsson}, {Eriksson},
  {J{\o}rgensen}, {Nordlund}, \& {Plez}}]{2008A&A...486..951G}
{Gustafsson}, B., {Edvardsson}, B., {Eriksson}, K., {J{\o}rgensen}, U.~G.,
  {Nordlund}, {\AA}., \& {Plez}, B. 2008, \aap, 486, 951. \eprint{0805.0554}

\bibitem[{{Haubois} et~al.(2009){Haubois}, {Perrin}, {Lacour}, {Verhoelst},
  {Meimon}, {Mugnier}, {Thi{\'e}baut}, {Berger}, {Ridgway}, {Monnier},
  {Millan-Gabet}, \& {Traub}}]{2009A&A...508..923H}
{Haubois}, X., {Perrin}, G., {Lacour}, S., {Verhoelst}, T., {Meimon}, S.,
  {Mugnier}, L., {Thi{\'e}baut}, E., {Berger}, J.~P., {Ridgway}, S.~T.,
  {Monnier}, J.~D., {Millan-Gabet}, R., \& {Traub}, W. 2009, \aap, 508, 923.
  \eprint{0910.4167}

\bibitem[{{Hauschildt} et~al.(1997){Hauschildt}, {Baron}, \&
  {Allard}}]{1997ApJ...483..390H}
{Hauschildt}, P.~H., {Baron}, E., \& {Allard}, F. 1997, \apj, 483, 390.
  \eprint{arXiv:astro-ph/9607087}

\bibitem[{{Iglesias} et~al.(1992){Iglesias}, {Rogers}, \&
  {Wilson}}]{1992ApJ...397..717I}
{Iglesias}, C.~A., {Rogers}, F.~J., \& {Wilson}, B.~G. 1992, \apj, 397, 717

\bibitem[{{Karovska} et~al.(1997){Karovska}, {Hack}, {Raymond}, \&
  {Guinan}}]{1997ApJ...482L.175K}
{Karovska}, M., {Hack}, W., {Raymond}, J., \& {Guinan}, E. 1997, \apjl, 482,
  L175

\bibitem[{{Lacour} et~al.(2009){Lacour}, {Thi{\'e}baut}, {Perrin}, {Meimon},
  {Haubois}, {Pedretti}, {Ridgway}, {Monnier}, {Berger}, {Schuller},
  {Woodruff}, {Poncelet}, {Le Coroller}, {Millan-Gabet}, {Lacasse}, \&
  {Traub}}]{2009ApJ...707..632L}
{Lacour}, S., {Thi{\'e}baut}, E., {Perrin}, G., {Meimon}, S., {Haubois}, X.,
  {Pedretti}, E., {Ridgway}, S.~T., {Monnier}, J.~D., {Berger}, J.~P.,
  {Schuller}, P.~A., {Woodruff}, H., {Poncelet}, A., {Le Coroller}, H.,
  {Millan-Gabet}, R., {Lacasse}, M., \& {Traub}, W. 2009, \apj, 707, 632.
  \eprint{0910.3869}

\bibitem[{{Ludwig} et~al.(1994){Ludwig}, {Jordan}, \&
  {Steffen}}]{1994A&A...284..105L}
{Ludwig}, H., {Jordan}, S., \& {Steffen}, M. 1994, \aap, 284, 105

\bibitem[{{Ohnaka}(2014)}]{2014A&A...568A..17O}
{Ohnaka}, K. 2014, \aap, 568, A17. \eprint{1407.0715}

\bibitem[{{Ohnaka} et~al.(2011){Ohnaka}, {Weigelt}, {Millour}, {Hofmann},
  {Driebe}, {Schertl}, {Chelli}, {Massi}, {Petrov}, \&
  {Stee}}]{2011A&A...529A.163O}
{Ohnaka}, K., {Weigelt}, G., {Millour}, F., {Hofmann}, K.-H., {Driebe}, T.,
  {Schertl}, D., {Chelli}, A., {Massi}, F., {Petrov}, R., \& {Stee}, P. 2011,
  \aap, 529, A163. \eprint{1104.0958}

\bibitem[{{Roe}(1986)}]{1986AnRFM..18..337R}
{Roe}, P.~L. 1986, Annual Review of Fluid Mechanics, 18, 337

\bibitem[{{Schwarzschild}(1975)}]{1975ApJ...195..137S}
{Schwarzschild}, M. 1975, \apj, 195, 137

\bibitem[{{V{\"o}gler} et~al.(2004){V{\"o}gler}, {Bruls}, \&
  {Sch{\"u}ssler}}]{2004A&A...421..741V}
{V{\"o}gler}, A., {Bruls}, J.~H.~M.~J., \& {Sch{\"u}ssler}, M. 2004, \aap, 421,
  741

\bibitem[{{Young} et~al.(2000){Young}, {Baldwin}, {Boysen}, {Haniff}, {Lawson},
  {Mackay}, {Pearson}, {Rogers}, {St.-Jacques}, {Warner}, {Wilson}, \&
  {Wilson}}]{2000MNRAS.315..635Y}
{Young}, J.~S., {Baldwin}, J.~E., {Boysen}, R.~C., {Haniff}, C.~A., {Lawson},
  P.~R., {Mackay}, C.~D., {Pearson}, D., {Rogers}, J., {St.-Jacques}, D.,
  {Warner}, P.~J., {Wilson}, D.~M.~A., \& {Wilson}, R.~W. 2000, \mnras, 315,
  635

\end{thebibliography}

\end{document}